\documentclass{article}
\usepackage[utf8]{inputenc}
\usepackage{epsfig}
\usepackage[font=small,labelfont=bf]{caption}
\usepackage{amssymb,amsmath,graphicx,latexsym}
 
\usepackage{graphicx}
\usepackage{float}
\usepackage{caption}
\usepackage{subcaption}
\usepackage{soul}
\usepackage{bm}
\usepackage[table]{xcolor}
\usepackage{url}
\usepackage{booktabs}
\usepackage{lmodern}
\usepackage[font=small,skip=0pt]{caption}
\usepackage[T1]{fontenc}
\usepackage{rotating}
\usepackage{pdflscape}
\usepackage{tabulary}
\usepackage{array}
\usepackage{booktabs}
\usepackage{appendix}
\usepackage{color, colortbl} % Highlighting the table rows/colomns
\usepackage[table]{xcolor}
\usepackage[first=0,last=9]{lcg}

\usepackage{graphicx}
\usepackage{caption}
\usepackage{geometry}
\usepackage{lipsum}
\makeatletter
\begin{document}

\begin{center}
\huge \textbf{\bf Copula-based Modeling for IBNR Claim Loss Reserving
}

\small
Samira Zaroudi $^a$, Mohammad Reza Faridrohani $^{b}$,Mohammad Hassan Behzadi$^{c}$,\\
and Hadi Safari-Katesari$^d$

% Author(s)

$^a$School of Mathematical and Statistical Sciences, Southern Illinois University Carbondale, IL 62901, USA\\
$^{b}$Department of Statistics, Faculty of Mathematical Sciences, Shahid Beheshti University, Tehran, Iran\\ % University/organization
$^c$Department of Statistics, Science and Research Islamic Azad University, Tehran, Iran\\
$^d$Department of Mathematical Sciences, Stevens Institute of Technology, Hoboken, NJ 07030, USA
\end{center}
%\end{minipage}
\vspace{1cm} % A bit of extra whitespace between the header and poster content

%----------------------------------------------------------------------------------------

%\begin{multicols}{4} % This is how many columns your poster will be broken into, a poster with many figures may benefit from less columns whereas a text-heavy poster benefits from more

%----------------------------------------------------------------------------------------
%	ABSTRACT
%----------------------------------------------------------------------------------------

\begin{abstract}
\noindent
%\textcolor{blue}{
There are growing concerns for reserves estimation of incurred but not reported (IBNR) claims in actuarial sciences. In this paper, we propose a copula-based dependency model to capture the relationship between two main IBNR reserve variables, i.e., the “time between two successive occurrences” and “delay time”. A maximum likelihood estimation method is used to estimate the parameters of the model.
%where the margins follow from exponential distributions.
A simulation study is conducted to evaluate the validity of the theoretical results. Moreover, the proposed method is applied to predict the number of claims for the next years of a portfolio from a major automobile insurer and is compared to the classical CL model forecasting. 
%}

Keywords: Copula; Event and report times; IBNR reserves; Run-off triangle; Third-party insurance.
\end{abstract}
\section{Introduction}
%\textcolor{blue}{
Reserves estimation for incurred but not reported (IBNR)  losses in the insurance policy period is one of the concerns of the actuarial profession.
%A major concern of insurers lies in calculating reserves for IBNR losses in the insurance policy period which is an inevitable phenomenon in insurance industry.
IBNR claims can stay open for a long period of time due to the juristic regulation processes and the size of claims.
%According to the size of claims and regulation processes for paying claims, loss claim can stay open for a long time. 
The difference between the time of occurrence and the time of payment by insurer
 will change the insurer's expected obligations and result to wrong amount of claim reserves. This is the place that the role of predicting and computing IBNR claim loss reserving is highlighted.
 %}
 %The classical method for estimating loss reserve contains data exploration from last periods and pattern recognition existing in a dataset to forecast future expected occurrences. 
% \textcolor{red}{
In this paper, we aim to estimate the number of IBNR claims by considering the dependency between the event time and report time of the losses for insurance companies. To do so, we use copula function to build the joint distribution of event and report times of the claims.
%}
%\textcolor{blue}{
 In order to compute the IBNR claim loss reserves, the classical methods apply data exploration to predict future expected losses,
 %In this way, different techniques have been proposed for computing IBNR claim loss reserving
 such as Bornhuetter-Ferguson (BF) method proposed by Bornhuetter and Ferguson [1], Benktander-Hovinen method proposed by Benktander [2] and Hovinen [3], Cape Cod method proposed by B\"{u}hlmann and Straub [4] and Stanard [5], and Chain-Ladder (CL) method proposed by Mack [6]. For more information about deficiency and properties of these methods, one can refer to [7-9]. 
 Another method to estimate the IBNR claim reserves is copula approach which models the dependency between occurring time and reporting time of an event.
 Copula is a powerful tool for modeling the dependency between different random variables. In statistical literature, there are many applications of copula such as [10] in aborts data, [11] in travels data, and [12] in biological networks. For insurance data, Pettere and Kollo [13] modeled the size of claims and delay time (between occurrence and report of the claims) by using Archimedean copula family. Zhao et al. [14] and Zhao and Zhou [15] presented a model for individual claims development by using semiparametric techniques of survival analysis and copula methods. Moreover, Shi and Freez [16] used a copula regression model to predict the unpaid losses to obtain the dependency between different lines of a business. Badescu et al. [17] showed that reported and IBNR claim processes are marked Cox processes, while Avanzi et al. [18] used Cox process method to predict the number of IBNR claims by using a dataset of Australian general insurer to model the reporting delay and risk exposure. Landriault et al. [19] computed the moments of total discounted IBNR claims by using a compound renewal process at a given time greater than zero. Also, they considered joint moments of total discounted IBNR claims and incurred and reported claims by using reporting lags and arrival times. Crevecoeur et al. [20] considered the problem of incurred but not yet reported (IBNYR) by using a granular method to model the time between occurrence and observation of claims. For more information about modeling IBNR and IBNYR claims, see [21-26].
\\
Another method to compute the insurance reserves for future obligations is multiplying the average of claims size to the average of claims number in each development time unit. The development time refers to the difference between the time that loss is occurred and the time that the loss is reported to the insurance company. In this paper, we estimate the number of IBNR claims through the following three steps:
\\
Step 1: Copula is applied to model the joint distribution of two marginal variables i.e., the “event time” and “report time” or equivalently “time between two successive occurrences” and “delay time”. 
%}
\\
Step 2: The individual conditional probability of reporting a claim happened in the development years is estimated based on Step 1 modeling.
\\
Step 3: The average claim size of a IBNR in the development years is estimated.
\\
Similar to Weissner [27], we assume that the marginal distributions i.e., “difference between two occurrences” and “delay time”, are exponential distributions with two different rates. We use copula to obtain the dependence between “difference between two successive occurrences” and “delay time”. This is while that Zhao and Zhou [15] applied copula approach to obtain the dependency between event time and report time.

The rest of the paper is organized as follows. Section \ref{sec2} reviews  CL method and copula model. Section \ref{sec3} specifies Clayton copula with event-report time variables as marginal distributions, and demonstrates estimation procedure of the IBNR claim numbers. Section \ref{sec4} conducts simulation study and real data application by using an automobile insurance dataset. Finally, Section \ref{sec5} concludes remarks.

%%%%%%%%%%%%%%%%%%%%%%%%%%%%
%%%%% Model Specification  %
%%%%%%%%%%%%%%%%%%%%%%%%%%%
\section{Model Specification}\label{sec2}
%\textcolor{blue}{
Copula is a tool to obtain the joint distribution of random variables, when the marginal distributions are available. It is also a strong technique to measure the size of both linear and nonlinear dependency between random variables.
Similar to Zhao and Zhou [15], we use copula approach to model the dependence structure of IBNR claim loss reserving but with different marginal distributions. Zhao and Zhou [15] applied copula approach to model the event and delay time for individual claim loss modeling. But, we use copula to obtain the joint distribution and  the dependence structure of the duration time between two successive events and the waiting time (reporting delay). 
%with illustrating our model by simulation study and real data application from a major insurer in automobile insurance market. 
In the Archimedean copula family, the Clayton copula [28] is the only absolutely continuous copula, which preserves the bivariate truncation. 
Oakes [29] applied Clayton model to obtain the joint distribution of the survival times, $T_{1}$ and $T_{2}$, which is interpreted as the ratio of the hazard rates of the conditional distribution of $T_{1}$ given $T_{2}=t_{2}$ to $T_{1}$ given $T_{2}>t_{2}$. 
In order to obtain the joint distribution and the dependence structure of the event and delay times to predict the number of IBNR claims, we propose a new dependence model via copula based on individual number of claims.
In our approach, the joint distribution of the marginal distributions i.e., the “difference between two successive occurrences” and “delay time”, are modeled by a parametric copula. Moreover, a Poisson process is fitted to the arrival process of claims. Similar to Jewell [30,31], the difference between the two successive occurrences and delays are fitted by using two exponential distributions. This model framework is more flexible than the competitive models for modeling IBNR claims. Moreover, we expect this framework generates more impressive and precise prediction for the number of IBNR claims. The evaluation of the accuracy of our framework is compared to the competitive models in the Section \ref{sec4}.
Here, we define the specification of our model framework and a traditional method for modeling the number of IBNR claim called CL method. First, we introduce the CL approach.
%}

\subsection{Chain-Ladder Method}
%\textcolor{blue}{
 Consider a portfolio of an automobile insurance company which is consist of $N>1$ run-off triangles of observations. Suppose that  $n$ ($\, 1\le n\le N$) indicates the number of portfolios (triangles), $i$ ($0\le i\le I$) shows the accident years (rows), and $j$ ($\, 0\le j\le J$) stands for the development years (columns). The number of claims in a portfolio with sample size $n$ for the accident year $i$ and development year $j$ is given by $X_{i,j}^{n}$ and the cumulative claims of the accident year $i$ up to the development year $j$ are denoted by 
 %}
\begin{equation}
  C_{i,j}^{n} =\sum _{k=0}^{j}X_{i,j}^{n},
\end{equation}
%\textcolor{blue}{
where $X_{i,j}^{n} =0$ for all $j>J$. The individual development factors for the accident year $i$ and development year $j$ are given as
%}
\begin{equation}\label{ff}
 f_{i,j}^{n} =\frac{\sum_{i=1}^{I-j}C_{i,j}^{n} }{\sum_{i=1}^{I-j}C_{i,j-1}^{n}},~~~  f_{i,j}^{n} =(f_{_{i,j} }^{1} ,\cdots ,f_{_{i,j} }^{N} )^{\top},
\end{equation}
\begin{equation}\label{eq3}
\widetilde{C}^{n} _{i,j} = C^{n}_{i,I-i} \prod _{j=I-i}^{J-1} f_{j}^{n}, 
\end{equation}
%\textcolor{blue}{
where $n\in \{1,\dots ,N\}, i\in \{1,\dots ,I\}$ and $j\in \{1,\dots ,J\}$, and $\widetilde{C} _{i,j}^{n} $ is the 
estimated number of IBNR reserve for the accident year $i$ and the development year \textit{j} [32]. Recently, the CL method faced high interest in insurance applications such as [33-36].
%}

\subsection{Copula Specification}
%\textcolor{blue}{
The concept of copula was introduced by Sklar theorem [37]. Nowadays, copula is a main technique to build the dependence structure for insurance and finance datasets. A copula $\mathcal{C}_\theta:[0,1]^n \xrightarrow{}[0,1]$ is a multivariate cumulative distribution function on $[0,1]\times[0,1]$  with marginal uniform distributions, where $\theta$ is an unknown dependence parameter of the copula. Sklar's Theorem states that any multivariate joint distribution can be written in terms of their univariate marginal distribution functions together with a copula. In the bivariate case, any joint distribution function $F_{T,S}$ corresponding to a bivariate random variable $(T,S)$ with univariate marginal distribution functions $F_T$ and $F_S$ can be obtained by 
%}
\begin{equation*}
    F_{t;s} (x,y)=\mathcal{C}_{\theta}(F_T(t),F_S(s)),
\end{equation*}
%\textcolor{blue}{
where $\mathcal{C}_{\theta}(\cdot)$ is the copula function with the dependence parameter $\theta$. One of the well-known class of copulas is Archimedean copulas. 
The advantage of Archimedean copula family is that the majority of copulas in this family have closed-form distribution functions. This is while that the copulas in the Gaussian copula family does not have closed-form distribution functions.
%In this family, we can obtain the most common Archimedean copulas explicitly, something not possible for instance in other class of copulas, e.g. in the Gaussian copula. 
Another characteristic of Archimedean copulas is that they allow to model the dependence structure of random variables in arbitrarily high dimensions with only one parameter. Here, we define Archimedean copulas. Let $\phi $ be a continuous and strictly decreasing function from $[0,1]$ to $[0,\infty]$  such that $\varphi \left(1\right)=0$. The pseudo-inverse of $\phi $ is the function $\phi^{[-1]} $  with domain $[0, \infty]$  and range $I=[0,1]$ which is given by
%}
\begin{align}
\phi^{[-1]}(z)=\left\{ \begin{array}{lr}
 \phi^{-1}(z) ,&\mbox{ $ 0\le z \le \phi(0)$} \\
0
, &\mbox{  $\phi(0) \le z \le \infty $}  \end{array} \right..
\end{align}
%\textcolor{blue}{
Notice that  $\phi^{[-1]}$ is continuous and non-increasing function on $[0,\infty]$, and strictly decreasing function on $[0,\phi(0)]$. Furthermore, we have $\phi^{[-1]}(\phi(u))=u $   on $I$, and
%}
\begin{eqnarray}
\phi(\phi^{[-1]}(z))=\left\{ \begin{array}{lr}
z ,&\mbox{ $ 0\le z \le \phi(0)$} \\
 \phi(0)
,&\mbox{  $\phi(0) \le z \le \infty $}  \end{array} \right.=min(z,\phi(0)).
\end{eqnarray}
Finally, if $\phi(0)= \infty $ then $\phi^{[-1]}=\phi^{-1} $ [28].
Let $\mathcal{C}$  be a copula function from $I^2$ to $I$  given by
\begin{equation}
 C_{\phi } (u;v)=\varphi ^{(-1)} \left(\varphi \left(u\right)+\varphi \left(v\right)\right).   
\end{equation}
It is easy to see that the copulas are invariant under monotone transformations of the marginal distribution. 
%\textcolor{blue}{
Therefore, monotone association measures such as copula-based Kendall's tau with the expression
%}
\begin{equation}
\tau =4\int _{\left[0,1\right]^{2} }C\left(u,v\right){\rm \; }dC\left(u,v\right)-1\in \left[-1,1\right.] 
\end{equation}
%\textcolor{blue}{
are used to obtain the size of dependency between marginal random variables [38]. This is while that the classical correlation measures such as Pearson's correlation coefficient only measures linear associations between marginal distributions. There are many studies to discuss how to select a copula for a given dataset, see [13,39]. The Clayton copula is an asymmetric Archimedean copula, which is able to measure positive dependency between random variables. It is also the most often applied and famous Archimedean copula in experimental applications [40].
%}
%, which has a constant time-dependence Oakes' cross-ratio function (Oakes, 1989). 
The Clayton copula function with association parameter $\theta$ is defined as 
\begin{align}
\mathcal{C}_{\theta}(w;t)=(t^{-\theta}+w^{-\theta}-1)^{-1/\theta},~~~~\theta\geq0.
\end{align}
Therefore, the joint density function of the Clayton copula is obtained as

\begin{equation} \label{2} 
c\left(t;w\right)=\left(\theta +1\right)\times \left(t\, \, w\right)^{-\left(\theta +1\right)} \times \left(t^{-\theta } +w^{-\theta } -1\right)^{-\left(2+\frac{1}{\theta } \right)},~~~~\theta\geq0.  
\end{equation} 
For further illustration and additional properties of the Clayton copulas, see [38,41,42]. Since the association parameter in Clayton copula only accepts positive values, this copula is convenient merely for positively associated random variables. 
%\textcolor{blue}{
When the dependence parameter $\theta$ converges to zero, Clayton copula demonstrates the independent between marginal random variables. The relationship between copula-based Kendall's $\tau$ correlation measure and Clayton copula is given as $\tau=\dfrac{\theta}{\theta +2}$, which enables us to measure the size of the copula-based Kendall's $\tau$ with known $\theta$. Moreover, the maximal value of $\tau$ is captured when $\theta$ goes to infinity. For more information, one can refer to [43-46].
%}

\section{Estimation of IBNR claim number}\label{sec3}
\subsection{The Number of IBNR Claim with Event-Report Time Modeling}
Let $T_i $ and $S_i $ denote the occurring time and the reporting time of an event, respectively. One can model the relationship between $T_i $ and $S_i $, directly, to predict the number of IBNR claims. Alternatively, we model this relationship indirectly according to the duration time between two successive events and the waiting time (reporting delay). Following Jewell [30,31], we assume that the positive random waiting times, denoted by  $W_i$’s, are independent and identically distributed (i.i.d) according to a common exponential distribution. We show the corresponding density of this distribution by $f_{W_{i} } (.|\beta _{2} )$, where $\beta_2$  is an unknown parameter. In our indirect method, the period of time for occurring the next event plays a pivotal role. We denote the duration time between two successive events by  $T^{*}$ that has exponential distribution with parameter $\beta_1$. The joint density function of $(T^{*} ,W)$  based on copulas is given as
\begin{equation}\label{3} 
f_{(T^{*} ,W)} \left(t,w|\beta _{1} ,\beta _{2} \right)=f_{T^{*} } \left(t|\beta 
_{1} \right) f_{W} \left(w|\beta _{2} \right) \; c\left[F_{T^{*} } \left(t|
\beta _{1} \right),F_{W} \left(w|\beta _{2} \right)\right],
\end{equation}
 where $c\left(t,w\right)=\frac{\partial ^{2} C\left({\rm t,s}\right)}{\partial t\partial 
w} $ is the density function of the copula C. Unfortunately, the recording of  $(T_{i}^{*} ,W_{i} )$'s are not possible, and so we cannot obtain the likelihood function of $(\beta_1,\beta_2)$ based on the joint density function defined in Eq. \eqref{3}. Instead, observations of the occurring event time $T_i$  and the reporting time $S_i$ are available. Therefore, we obtain the joint density function of $(T_{i} ,{\rm \; }S_{i} )$  by using the joint density function of $(T^{*}_{i} ,W_{i} )$  represented in Eq. \eqref{3}.

 Notice that the occurrence time of the $i^{th}$ event, $T_i$, is obtained by summing over all duration times between two successive events up to that time, i.e.,  $T_{i} =T_{1}^{*} +T_{2}^{*} +\cdots +T_{i}^{*} =T_{i-1} 
+T_{i}^{*}$. Then, $T_i$ has the Gamma distribution   $\Gamma (i,\beta _{1})$, because $T^{*}_i$'s are iid and follow exponential distribution. On the other hand, it is easy to see that $S_{i} =T_{i} +W_{i} =T_{i-1} +T_{i}^{{
\rm *}} +W_{i} $. Therefore, the joint density function of $(T_{i},S_{i} )$ is obtained as

\begin{align}\label{ftisi}
\begin{split}
f_{(T_{i} ,S_{i} )}(t,s) =&f_{(T_{i} ,W_{i} )} (t,s-t)=f_{(T^{*}_{i}+T_{i-1},W_{i})} (t,s-t)\\
=& \int _{0}^{t}f_{(T_{i}^{*},T_{i-1},W_{i} )} (t-u, u,s-t)du\\
=& \int _{0}^{t}f_{(T_{i}^{*},W_{i}|T_{i-1})} (t-u,s-t|u) f_{T_{i-1}}(u)du\\
=&\int _{0}^{t}f_{(T_{i}^{*},W_{i})} (t-u,s-t)f_{T_{i-1}}(u)du,
\end{split}
\end{align}
where the $(i-1)^{th} $event time, $T_{i-1}$, is independent from  $(T_{i}^{*},W_{i})$ and has Gamma distribution $\Gamma (i-1,\beta _{1}).$
Moreover, the joint distribution between $T^{*}_{i}$ and $W_{i}$ is obtained by using the Clayton copula defined in Eq. \eqref{2} as follows
\begin{align}
\begin{split}
f_{(T_{i},S_{i} )}(t,s)=&\int _{0}^{t}\frac{e^{-(t-u)
\beta _{1} } (t-u)^{(i-2)}\beta_{1}^{(i-1)}}{\Gamma(i-1
)})\beta_{1} \beta _{2} (\theta +1) \\ 
&\times e^{-(\beta_{1} u)}e^{-\beta_{2}(s-u)}((1-e^{-(\beta_{1} u)})(1-e^{-\beta _{2}(s-u)}))^{-(\theta+1)}  \\ 
&\times((1-e^{-(\beta_{1}u)})^{-\theta}+(1-e^{-\beta_{2}(s-u)})^{-\theta}-1)^{-(2+\frac{1}{\theta})}du. 
\end{split}
\end{align}
Then, the likelihood function of $(\beta_1,\beta_2)$ based on $(T_i,S_i)$ is as follows
\begin{align}\label{like}
\begin{split}
L(\beta_1,\beta_2,\theta;(t_1,s_1),\cdots,(t_n,s_n))
=& \prod^{n}_{i=1} f_{(T_{i},S_{i} )}(t_{i},s_{i} )\\
=&\prod^{n}_{i=1} \int_{0}^{t_i}\frac{e^{-(t_i-u)\beta_{1}} (t_i-u)^{(i-2)}\beta_{1}^{(i-1)}}{\Gamma(i-1)})\beta_{1}\beta_{2} (\theta+1)\\  
&\times e^{-(\beta_{1}u)}e^{-\beta_{2}(s_i-u)}((1-e^{-(\beta_{1}u)} )(1-e^{-\beta_{2}(s_i-u)} ))^{-(\theta 
+1)}\\ 
&\times ((1-e^{-(\beta _{1}u)} )^{-\theta 
} +(1-e^{-\beta_{2}(s_i-u)})^{-\theta}-1)^{-(2+\frac{1}{\theta})}du.
\end{split}
\end{align}

The maximum likelihood estimation (MLE) of $\beta _{1}$, $\beta _{2}$, and $\theta $ can be obtained by maximizing the likelihood function in Eq. \eqref{like}.
 
\subsection{Delay probability }
After estimating the joint density function of  $f_{(T_{i},S_{i})}(t,s)$ defined in Eq. \eqref{ftisi}, we are able to predict the number of claims reported in the next years. By using the information about $i^{th}$  event occurrences in the $j^{th}$  year, we can estimate the probability of reporting this event in the next $(i+j)^{th}$  years as follows

\begin{align}
\hat{p}_{i,j}^{(l)} =\hat{P}(S_{i}\in I_{j+l}|T_{i}\in I_{j})
=\frac{\hat{P}(T_{i} \in I_{j},S_{i} \in I_{j+l})}{\hat{P}(T_{i} \in I_{j})},~~ l=1, \dots , n_J-j,
\end{align} 
where $n_{J}$ is the upper bound of delay time. 

%%%%%%%%%%%%%%%%%%%%%%%%
%%% IBNR claim number estimation
%%%%%%%%%%%%%%%%%%%%%%%%
\subsection{IBNR claim number estimation }
%\textcolor{blue}{
In order to estimate the number of IBNR claims, we need to obtain $\hat{N}^{l}_{j}$, which is the expected number of occurrences related to the reporting the event in the next $(j+l)^{th}$ years for $j=1,\cdots, n_J$. Therefore, it can predict the number of claims incurred in the year $(j+l)$.
%}
Hence, one needs to estimate the expected number of IBNR claims by using following  equation
\begin{equation} \label{5}
\hat{N}_{i,j}^{l} =\sum_{k=1}^{n_{i}}  \hat{p}_{k,j}^{(l)}, ~~~i=1,\cdots, n_I. 
\end{equation} 
 %Individual claim loss models have been investigated under this framework.
%------------------------------------------------
%%%%%%%%%%%%%%%%%%%%%%%%
%%% Data Analysis
%%%%%%%%%%%%%%%%%%%%%%%%
\section{Data Analysis}\label{sec4}
%\textcolor{blue}{
In this section, we apply the proposed methods in Section \ref{sec3} in simulation study and a real dataset. We conduct comparison study to compare the proposed methods with the competitor methods. Moreover, the performance of the maximum likelihood estimator of $(\beta_1,\beta_2,\theta)$ defined in Eq. \eqref{like} is considered. By using the estimator introduced in Eq. \eqref{5}, we predict the claim number in the next years in a third-party insurance policy of an insurance company in Iran.
%}
The performance of the proposed model is compared with the CL model forecasting. 
%%%%%%%%%%%%%%%%%%%%%%%%
%%% Simulation Study
%%%%%%%%%%%%%%%%%%%%%%%%
\subsection{Simulation Study}
As mentioned in section \ref{sec3}, $T^{*}_i$'s and $W_i$’s are dependent random variables and have exponential distributions with different rate parameters. 
%\textcolor{blue}{
In order to generate a sequence of dependent observations $t^*_i$ and  $w_i$ from random variables $T^{*}_i$ and $W_i$, respectively, we apply accept-reject algorithm as follows.
Let $Y_i=f_{W_i|T^*_i=t^*}(w|t^*)$ and $V=f_{W_i}(w)\sim exp(\beta_1)$, where $f_{Y_i}$ and $f_{W_i}$ have common support with $M=\sup f_{Y_i}/f_{W_i}<\infty$. consider $Y\sim f_{Y_i}$. Then, 
%}
\begin{itemize}
    \item [a)]
generate $U\sim uniform(0,1)$ and $V=f_{W_i}(w)$ independently,
\item [b)]
if $U< \dfrac{1}{M} f_Y(V)/f_V(V)$, set $Y=V$; otherwise, return to step a).
\end{itemize}
Here, our goal is to generate the data from $W_i$ which are dependent of $T^*_i$. 
%\textcolor{blue}{
That is we have $Y_i=f_{W_i|T^*_i=t^*}(w|t^*)$. The simulated datasets are generated by using the accept-reject algorithm to be used to estimate different parameters of the model, i.e., $\beta_1$, $\beta_2$, and $\theta$. The MLE of the parameters are conducted for different sample sizes, i.e., 50, 150 and 200, where the number of replication is 100,000.
Moreover, the initial values of the scale parameters for the MLE algorithm are considered as the mean of random sample. For determining the initial values for $\theta$, we computed the Kendall's tau ($\hat{\tau}$) for generated samples and obtained the initial value of $\theta$ by using  $\theta=2\tau/(1-\tau)$.
The mean of the MLEs, mean square errors, and  bias of the estimated parameters are reported in Table 1.
%}
%The replication number in our algorithm is selected as 100,000 repetitions. 
Note that in this simulation, we selected the real parameters as $\beta_1=0.5, \beta_2=0.5, \theta=1.5$. 
%\textcolor{blue}{
Table 1 demonstrates the average of MLE’s, their mean squared error (MSE)’s and biases for parameters $\beta_{1}$, $\beta_{2}$, and $\theta$ with real values 0.5 , 
0.5, and 1.5, respectively.
\\
In Table 2, the ratios of the simulated number of claims reported in a typical year, i.e., 2016, but occurred over the past 7 years, i.e., during 2010-2016, are reported.
%}
%%%%%%%%%%%%%%%%%%%%%%%%%%%%%%%%%%%%%%%%%%%%%%%%%%%
%%%%%%%%%%%%%%%%%%%%%%%
%%%%%%%%%%%%%%%%%%%%%%%%%%%%%%%%%%
%%% Real Data Application
%%%%%%%%%%%%%%%%%%%%%%%%%%%%%%%%
\subsection{Real Data Application}
%\textcolor{blue}{
In this section, we apply our proposed copula model and CL method to a real dataset from a major automobile insurer in Iran. 
%%%%%%%%%%%%%%%%%%%%%%%%%%%%%%%%%%
%%%%%%%%%%%%%%%%%%%%%%%%%%%%
In particular, we used the observations of a subsample of 140,228 policies recorded in the portfolio of the insurance company during 7 years from 2010 to 2016. We fitted the exponential distribution to marginal distributions, i.e., the ``duration time between two successive events'' and the ``reporting delay time'' in our dataset. We carried out Kolmogorov-Smirnov test in which
%for test of exponentiality on duration time between two successive events and test of exponentiality on reporting delay time 
 p-values are 0.141 and 0.214, respectively.
Therefor, we can assume that the marginal distributions of our copula model are following exponential distributions.
As mentioned in Section \ref{sec3}, we estimate all parameters using the MLE method. Notice that we provided Tables \ref{table1}-\ref{table8} in the Appendix.
%Moreover, we follow the date of occurrence $(T)$, report time $(S)$, reporting delay time $(W)$ between date of occurrence and report time, and distance between occurrences.
%Table 2 provides a summary of the frequency statistics for delay time between occurrence and report. As indicated in Table 2, the most percentage is related to the base year or final year which report time and date of occurrences are equal. The Tables are provided in the Appendix.
%%%%%%%%%%%%%%%%%%%%%%%%%%
%In order to compare our copula method to the competitive method, CL,
First, we apply CL method to this dataset.
The upper triangle of Table \ref{table3} provided the real number of cumulative claims and the lower triangle of this table demonstrated the estimated number of cumulative claims based on the CL method for the years between 2010 and 2015. 
In Table \ref{table3}, first, we obtained the number of claims in each development year for different accident years by using Eq. \eqref{eq3}. Then, we obtained the number of cumulative claims.
The development year refers to the difference between the year that loss is occurred and the year that the loss is reported to the insurance company. For example, the development year equal zero means that the occurrence time and reporting time of the losses are in the same year and the development year equal 3 means that the losses are reported 3 years after occurrence of the loss. Also, $f_{i,j}^{n}$ is the individual development factors for the accident year $i$ and development year $j$ defined in Eq. \eqref{ff}. Similarly, we provided the predicted number of cumulative claims based on the CL method for the years 2010-2016 in Table \ref{table4}. Now, we apply copula method to this dataset. We provided the estimated number of cumulative claims based on the copula method for the years between 2010 and 2015
in Table \ref{table6}, and for the years 2010 to 2016 in Table \ref{table7}. We obtained the number of claims in each development year for different accident years by using Eq. \eqref{5}. 
\\
In order to compare the performance of our proposed copula model and CL method in predicting the number of reported claims during different development years, we provided the percentage of the proportional absolute value of errors based on CL method for the years 2010-2015 in Table \ref{table5} and based on copula model for  the years 2010-2015 in Table \ref{table8}. The percentage of the proportional absolute value of errors in Tables \ref{table5} is computed by subtracting the values of Tables \ref{table4} from corresponding values of Table \ref{table3}, which result is divided to the corresponding values of Table \ref{table3}. Similarly, The percentage of the proportional absolute value of errors in Tables \ref{table8} is computed by subtracting the values of Tables \ref{table7} from corresponding values of Table \ref{table6}, which result is divided to the corresponding values of Table \ref{table6}. Obviously, there is not any error value for the year 2016 in Tables \ref{table5} and \ref{table8}. 
%}
%The results is shown in Table \ref{table5} for Cl method. Also the same procedure Table \ref{table8} for copula method. 
%\textcolor{red}{
For more illustration, we provided an example, which shows how to compute the error values in Tables \ref{table5} and \ref{table8}. The predicted number of claims based on CL method in Table \ref{table3} for accident year 2015 and development year $1$ is $23719$. This is while that the real value of the number of claims in Table \ref{table4} is $22769$. The percentage of the proportional absolute value of error based on CL method in Table \ref{table7} is equal to $|22769-23719|\times 100 /23719=4.0052$. The corresponding percentage of the proportional absolute value of error based on copula method in Table \ref{table8} for accident year 2015 and development year $1$ is obtained as $ |22769-22779|\times 100 /22779=0.0439$. Therefor, the percentage of the proportional absolute value of error based on copula method $(0.0439)$ is smaller than the error term based on CL method ${4.0052}$. Similarly, we can obtain all percentage of the proportional absolute value of error in Table \ref{table5} and Table \ref{table8}.
By comparing the results of the percentage of the proportional absolute value of errors based on CL method in Table \ref{table5} and copula method in Table \ref{table8}, we can conclude that our proposed copula method is performing better than CL method.
%}
%The corresponding results for the CL method are calculated and the forecasting numbers of cumulative losses during (2010-2016) are presented in Table 4. The same results for proposed copula method are given in Table 7.
%Table \ref{table3} provided the absolute value of the relative error between real number of losses and the predicted losses for the years 2010-2015.
%the forecasting cumulative losses for six development years (2010-2015), under CL method and our proposed copula model are reported in Table 5 and 8, respectively. 
%Table 6 and Table 9 show the corresponding proportional errors of forecasting claim numbers (in percent) for 2010-2015 years, under CL method and the proposed copula model. As an example, forecasting loss claim numbers in Table 4 for development year 4 in the year 2014 is 23,160 and in Table 5 is 23,367. So, the proportional rate of forecasting loss claim for this year is 0.89\% that is illustrated in Table 6.
%The comparison of the error rates represented in Table \ref{table5} and \ref{table8}, demonstrates that the performance of our proposed copula-based method has superior to the CL method. 
%%%%%%%%%%%%%%%%%%%%%%%%%%%%%%%%%%
%%% Conclusions
%%%%%%%%%%%%%%%%%%%%%%%%%%%%%%%%
\section{Conclusions}\label{sec5}
%\textcolor{blue}{
In this paper, we proposed a copula method to predict the IBNR claims. To do so, we applied a well-known family of copulas called Archimedean family. Particularly, we used Clayton copula to find the joint distribution between
%The main goal of this paper is to extend the literature in IBNR claim loss reserving methods. We reviewed the existing individual claim loss reserving models, and then, we proposed a model according to 
``difference between two occurrences'' and ``delay time''.
%by applying Clayton copula. 
%We used Clayton copula to jointly model the difference between two occurrences and the delay variables. 
In order to assess the performance of the proposed method, we applied a well-known and competitive CL method and compared the results through simulation and real data application.
The simulation study indicates that the proposed procedure can produce efficient estimates and improve predictions for the event delay numbers for the next year. Moreover, we used an empirical observation dataset from an insurance portfolio of a major automobile insurer in Iran. The results indicated that the performance of our proposed copula-based method has superior to CL method.
As future directions, our method can be extended to the case that the actual event times are forgotten. Moreover, one can extend this method to the non-exponential marginal distributions. 
%}

%\newpage
%%%%%%%%%%%%%%%%%%%%%%%%%%%%%%%%%%%%%%%%%%%%%%%%%%
%\section*{Appendix}

\newpage
\section*{Appendix}

%%%%%%%%%%%%%%%%%%%%%%%%%%%%%%%%
\begin {table}[!htbp]
\caption{The average of MLE (Maximum Likelihood Estimation), MSE (Mean Squared Error), and biases for parameters 
($\beta_{1},\beta_{2},\theta $) with real values ($\beta_{1}=0.5, \beta_{2}=0.5,\theta =1.5$) for sample size $n=50,150,$ and $250$.}
\label{table1}
\begin{center}
\begin{tabular}{c c c c c c c c c c c c c c c }
\hline 
    && \multicolumn{3}{c}{MLE} && Kendall's tau && \multicolumn{3}{c}{MSE} && \multicolumn{3}{c}{Bias} \\ 
    \cline{3-5}   \cline{7-7}  \cline{9-11}  \cline{13-15}
n   && $\hat{\beta}_{1} $ & $\hat{\beta}_{2} $ & $\hat{\theta}$ && $\tau$ && $\beta_{1}$ & $\beta_{2}$ & $\theta$ && $\beta_{1}$ & $\beta_{2}$ & $\theta$ \\ 
\hline 
50  && 0.570 & 0.587 & 1.746 && 0.466 && 0.051 & 0.063 & 0.097 && 0.07 & 0.087 & 0.246 \\ %\hline 
150 && 0.545 & 0.561 & 1.675 && 0.456 && 0.009 & 0.006 & 0.013 && 0.045 & 0.061 & 0.175 \\ %\hline 
200 && 0.507 & 0.523 & 1.537 && 0.434 && 0.004 & 0.008 & 0.009 && 0.007 & 0.023 & 0.037 \\ 
\hline
\end{tabular}
\end{center}
\end{table}
%%%%%%%%%%%%%%%%%%%%%%%%%
%%%%%%%%%%%%%%%%%%%%%%%%%%%%%%
\begin{table}[!htbp]
\caption{Simulation results for the ratio of the number of claims reported in  the year 2016 to the number of claims occurred over the years 2010-2016 with different sample sizes ($n= 50, 150,$ and $200$).}\label{table2}
\begin{center}
\begin{tabular}{c c c c c c c c c} \hline 
\multicolumn{9}{c}{Ratios} \\ 
\hline
&\multicolumn{8}{c}{Years} \\ 
%\hline
    \cline{3-9} 
 %\backslashbox{ n}{Year}
%\textbf{n} \| \textbf{Year}
n  && 2016 & 2015 & 2014 & 2013 & 2012 & 2011 & 2010 \\ 
\hline 
50 && 0.6400 & 0.2200 & 0.0600 & 0.0400 & 0.0200 & 0.0200 & 0.00 \\ 
%\hline 
150 && 0.6933 & 0.1400 & 0.0733 & 0.0600 & 0.0200 & 0.0067 & 0.0067 \\ 
%\hline 
200 && 0.7400 & 0.1600 & 0.0700 & 0.0150 & 0.000 & 0.0050 & 0.0050 \\ 
\hline 
\end{tabular}
\end{center}
\end{table}
%%%%%%%%%%%%%%%%%%%%%%%%%%%%%%%%%%%%%%%%
%%%%%%%%%%%%%%%%%%%%%%%%%%%%%%%%%%%%%%%%%%%%%%%%
\begin {table}[H]
\caption{Estimated numbers of cumulative claims based on CL method for the years 2010-2015}\label{table3}
\begin{center}
\begin{tabular}{c c c c c c c} \hline 
\multicolumn{7}{c}{Development year} \\ 
\hline 
Accident year & 0 & 1 & 2 & 3 & 4 & 5 \\ 
\hline 
2010 & 5,866 & 9,237 & 9,720 & 9,785 & 9,805 & 9,810 \\ 
%\hline 
2011 & 19,295 & 23,307 & 23,897 & 24,067 & 24,113 & \textbf{24,125} \\ 
%\hline 
2012 & 20,987 & 25,298 & 25,978 & 26,117 & \textbf{26,168} & \textbf{26,181} \\ 
%\hline 
2013 & 18,923 & 22,757 & 23,281 & \textbf{23,427} & \textbf{23,473} & \textbf{23,485} \\ 
%\hline 
2014 & 18,977 & 22,539 & \textbf{23,176} & \textbf{23,321} & \textbf{23,367} & \textbf{23,379} \\ 
%\hline 
2015 & 19,329 & \textbf{23,719} & \textbf{24,389} & \textbf{24,542} & \textbf{24,590} & \textbf{24,603} \\ 
\hline 
$f_{i,j}^{n}$ &  & 1.227132 & 1.028251 & 1.006276 & 1.001950 & 1.000510 \\ 
\hline
\end{tabular}
\end{center}
\end{table}

%%%%%%%%%%%%%%%%%%%

\begin {table}[!htbp]
\begin{center}
\caption{Estimated number of cumulative claims based on the CL method for the years 2010-2016.}
\label{table4}
\begin{tabular}{c c c c c c c c} \hline 
\multicolumn{8}{c}{Development year} \\ \hline 
Accident year & 0 & 1 & 2 & 3 & 4 & 5 & 6 \\ 
\hline 
2010 & 5866 & 9237 & 9720 & 9785 & 9805 & 9810 & 9813 \\ 
%\hline 
2011 & 19295 & 23307 & 23897 & 24067 & 24113 & 24131 & \textbf{24138} \\ 
%\hline 
2012 & 20987 & 25298 & 25978 & 26117 & 26174 & \textbf{26192} & \textbf{26206} \\ 
%\hline 
2013 & 18923 & 22757 & 23281 & 23397 & \textbf{23445} & \textbf{23469} & \textbf{23484} \\ 
%\hline 
2014 & 18977 & 22539 & 22977 & \textbf{23113} & \textbf{23191} & \textbf{23213} & \textbf{23223} \\ 
%\hline 
2015 & 19329 & 22769 & \textbf{23368} & \textbf{23517} & \textbf{23605} & \textbf{23634} & \textbf{23655} \\ 
%\hline
2016 & 10946 & \textbf{13332} & \textbf{13683} & \textbf{13763} & \textbf{13792} & \textbf{13801} & \textbf{13805} \\ 
\hline 
$f_{i,j}^{n}$ &  & 1.21794 & 1.02632 & 1.00591 & 1.00205 & 1.00068 & 1.00031 \\ 
\hline
\end{tabular}
\end{center}
\end{table}

%%%%%%%%%%%%%%%%%%%%%%%%%%%%%%%

%\begin {table}[H]\label{table7}
%\caption{Estimated number of losses based on the proposed copula model for the calendar years 2010-2016.}
%\begin{center}
%\begin{tabular}{c c c c c c c c}
%\hline 
%\multicolumn{8}{c}{Development year} \\ \hline 
%Accident year & 0 & 1 & 2 & 3 & 4 & 5 & 6 \\ 
%\hline 
%2010 & 5866 & 3371 & 483 & 65 & 20 & 5 & 3 \\ 
%\hline 
%2011 & 19295 & 4012 & 590 & 170 & 46 & 18 & \textbf{\textit{6}} \\ 
%\hline 
%2012 & 20987 & 4311 & 680 & 139 & 57 & \textbf{\textit{24}} & \textbf{8} \\ 
%\hline 
%2013 & 18923 & 3834 & 524 & 116 & \textbf{\textit{56}} & \textbf{24} & \textbf{3} \\ 
%\hline 
%2014 & 18977 & 3562 & 438 & \textbf{\textit{167}} & \textbf{53} & \textbf{19} & \textbf{7} \\ 
%\hline 
%2015 & 19329 & 3440 & \textbf{\textit{610}} & \textbf{178} & \textbf{61} & \textbf{30} & \textbf{5} \\ 
%\hline 
%2016 & 10946 & \textbf{\textit{4150}} & \textbf{581} & \textbf{156} & \textbf{54} & \textbf{27} & \textbf{4} \\
%\hline
%\end{tabular}
%\end{center}
%\end{table}
%%%%%%%%%%%%%%%%%

%%%%%%%%%%%%%%%%%%%%%%%%%%%%%%%%%%%%%%%%%%%%%%%%%%%%%%%%
%%%%%%%%%%%%%%%%%%%%%%%%%%%%%%%%%%%%%%%%%%%%%%%%%%%%%%
\begin {table}[!htbp]
\caption{The percentage of the proportional absolute value errors of number of claims based on CL method in compared with the values presented in Table \ref{table4}}
\label{table5}
\begin{center}
\begin{tabular}{c c c c c c c}  
\hline 
\multicolumn{7}{c}{Development year} \\ 
\hline 
Accident year & 0 & 1 & 2 & 3 & 4 & 5 \\ 
\hline 
2010 & - & - & - & - & - & - \\ 
%\hline 
2011 & - & - & - & - & - &  0.0249  \\ 
%\hline 
2012 & - & - & - & - &  0.0229  &  0.0420  \\ 
%\hline 
2013 & - & - & - &  0.8586 &  0.1193
 &  0.0681
\\ 
%\hline 
2014 & - & - & 0.8586 & 0.8919& 0.7532& 0.7100
\\ 
%\hline 
2015 & - & 4.0052 &4.1863 &  4.1765& 4.0057 &3.9385
\\  
\hline
\end{tabular}
\end{center}
\end{table}

%%%%%%%%%%%%%%%%%%%%%%%%%%%%%%%%%%%%%%

%%%%%%%%%%%%%%%%%%%%%%%%%%%%%%%%%%%%%%%%%%%%%%%%%%%%%%
\begin {table}[H]
\caption {Estimated number of cumulative claims based on the copula method for the years 2010-2015}\label{table6}
\begin{center}
\begin{tabular}{c c c c c c c} 
\hline 
\multicolumn{7}{c}{Development year} \\ 
\hline 
Accident year & 0 & 1 & 2 & 3 & 4 & 5 \\ 
\hline 
2010 & 5866 & 9237 & 9720 & 9785 & 9805 & 9810 \\ 
%\hline 
2011 & 19295 & 23307 & 23897 & 24067 & 24113 & \textbf{24140} \\ 
%\hline 
2012 & 20987 & 25298 & 25978 & 26117 & \textbf{26177} & \textbf{26189} \\ 
%\hline 
2013 & 18923 & 22757 & 23281 & \textbf{23386} & \textbf{23465} & \textbf{23484} \\ 
%\hline 
2014 & 18977 & 22539 & \textbf{22981} & \textbf{23143} & \textbf{23210} & \textbf{23234} \\ 
%\hline 
2015 & 19329 & \textbf{22779} & \textbf{23389} & \textbf{23550} & \textbf{23620} & \textbf{23651} \\ 
\hline
\end{tabular}
\end{center}
\end{table}

%%%%%%%%%%%%%%%

\begin {table}[!htbp]
\caption{Estimated number of cumulative claims based on the copula method for the years 2010-2016.}
\label{table7}
\begin{center}
\begin{tabular}{c c c c c c c c} 
\hline 
\multicolumn{8}{c}{Development year} \\ 
\hline 
Accident year & 0 & 1 & 2 & 3 & 4 & 5 & 6 \\ 
\hline 
2010 & 5866 & 9237 & 9720 & 9785 & 9805 & 9810 & 9813 \\ 
%\hline 
2011 & 19295 & 23307 & 23897 & 24067 & 24113 & 24131 & \textbf{24137} \\ 
%\hline 
2012 & 20987 & 25298 & 25978 & 26117 & 26174 & \textbf{26198} & \textbf{26206} \\ 
%\hline 
2013 & 18923 & 22757 & 23281 & 23397 & \textbf{23453} & \textbf{23477} & \textbf{23480} \\ 
%\hline 
2014 & 18977 & 22539 & 22977 & \textbf{23144} & \textbf{23197} & \textbf{23216} & \textbf{23223} \\ %\hline 
2015 & 19329 & 22769 & \textbf{23379} & \textbf{23557} & \textbf{23618} & \textbf{23648} & \textbf{23653} \\ 
%\hline 
2016 & 10946 & \textbf{15096} & \textbf{15677} & \textbf{15833} & \textbf{15887} & \textbf{15914} & \textbf{15918} \\
\hline
\end{tabular}
\end{center}
\end{table}
%%%%%%%%%%%%%%%%%%%%%%%
\begin {table}[!htbp]
\begin{center}
\caption {The percentage of the proportional absolute value errors of number of claims based on copula method in compared with the values presented in Table \ref{table7}.}
\label{table8}
\begin{tabular}{c c c c c c c} 
\hline 
\multicolumn{7}{c}{Development year} \\ 
\hline 
Accident year & 0 & 1 & 2 & 3 & 4 & 5 \\ 
\hline 
2010 & - & - & - & - & - & - \\ 
%\hline 
2011 & - & - & - & - & - & 0.0373 \\ 
%\hline 
2012 & - & - & - & - & 0.0115&	0.0344
\\ 
%\hline 
2013 & - & - & - & 0.0470&	0.0511&	0.0298
 \\ 
%\hline 
2014 & - & - &0.0174&	0.0043&	0.0560	&0.0775 \\ 
%\hline 
2015 & - 	&0.0439&	0.0428	&0.0297	&0.0085& 0.0127\\
\hline
\end{tabular}
\end{center}
\end{table}
%%%%%%%%%%%%%%%%%%%%%%%%%%%%%%%%%%%%%%%%%%%%%%

\end{document}